\documentclass[preprint,showpacs,preprintnumbers,amsmath,amssymb]{revtex4}
\usepackage{graphicx}
\usepackage{dcolumn}
\usepackage{bm}

\newcommand\br{\begin{eqnarray}}
\newcommand\er{\end{eqnarray}}
\newcommand\be{\begin{equation}}
\newcommand\ee{\end{equation}}

\newcommand\bc{\begin{center}}
\newcommand\ec{\end{center}}

\newcommand\NPB[3]{\textsl{Nucl. Phys.} \textbf{B#1}, #3 (#2)}

\newcommand\PRD[3]{\textsl{Phys. Rev.} \textbf{D#1}, #3 (#2)}

\newcommand\PLB[3]{\textsl{Phys. Lett.} \textbf{#1B}, #3 (#2)}
\newcommand\CQG[3]{\textsl{Class. Quantum Grav.} \textbf{#1}, #3 (#2)}

\newcommand\AoP[3]{\textsl{Ann. of Phys.} \textbf{#1}, #3 (#2)}
\newcommand\RMP[3]{\textsl{Rev. Mod. Phys.} \textbf{#1}, #3 (#2)}

\newcommand\IJMPA[3]{\textsl{Int. J. Mod. Phys.} \textbf{A#1}, #3 (#2)}

\newcommand\JPA[3]{\textsl{J. Physics} \textbf{A#1}, #3 (#2)}

\newcommand\MPLA[3]{\textsl{Mod. Phys. Lett.} \textbf{A#1}, #3 (#2)}

\begin{document}


\title{Bags and Confinement Governed by S.S.B. of Scale Invariance}


\author{E. I. Guendelman}
\email{guendel@bgu.ac.il} \affiliation{ Physics Department, Ben
Gurion University of the Negev, Beer Sheva 84105, Israel}

\bigskip

\begin{abstract}
A general coordinate invariant theory is constructed where confinement of gauge fields and gauge dynamics in general is
governed by the spontaneous symmetry breaking (s.s.b.) of scale invariance. The model uses two measures of integration in the action,
the standard $\sqrt{-g}$ where $g$ is the determinant of the metric and another measure $\Phi$ independent
of the metric. To implement scale invariance (S.I.), a dilaton  field is introduced. Using the first order formalism, curvature ( $\Phi R$ and  $\sqrt{-g}R^{2}$ ) terms , gauge field term( $\Phi\sqrt { - F_{\mu \nu }^{a}\, F^{a}_{\alpha\beta}g^{\mu\alpha}g^{\nu\beta}}$ and  $\sqrt{-g} F_{\mu \nu }^{a}\, F^{a}_{\alpha\beta}g^{\mu\alpha}g^{\nu\beta}$ ) and dilaton kinetic 
  terms are introduced in a conformally invariant way. Exponential potentials for the dilaton break down (softly) the conformal invariance down to global scale invariance, which also suffers s.s.b. after integrating the equations of motion. The model has a well defined flat space limit.  As a result of the s.s.b. of scale invariance phases with different vacuum energy density appear. Inside the bags, that is in the regions of larger vacuum energy density, the gauge dynamics is normal, that is non confining, while for the region of smaller vacuum energy density, the gauge field dynamics is confining.  Likewise, the dynamics of scalars, like would be Goldstone bosons, is suppressed inside the bags.
\end{abstract}


\maketitle
\section{Introduction}
In the bag model of confinement \cite{MIT} a very intriging connection is made between
the non zero value of the vacuum energy density (the Bag constant) and the almost free non confining dynamics for 
the gauge fields that holds inside the bags. Gauge fields are prevented to flow to the region of lower vacuum energy density
by the M.I.T. bag model boundary conditions. The lower vacuum energy density state represents the "confinement region" then.

On the other hand, in modern cosmology vacuum energy density is a central theme. First in the formulation of the inflationary phase of the early universe \cite{Inflation}, which provides an attractive scenario for solving
 some of the fundamental puzzles of the standard Big Bang model,
like the horizon and the flatness problems as well as providing a
framework for sensible calculations of primordial density perturbations. Then more recently, we had also evidence for the accelerated expansion
of our present universe \cite{Accelerated}, which could be also explained by some vacuum energy density in our present universe, this present vacuum energy density 
being of course of much smaller magnitude than that of the one postulated for the inflationary phase of the early universe.

Also in the context of modern cosmology, as well as for the bag model, we need two phases. In cosmology the two phases they should be connected
through cosmological evolution, while in the bag model through the boundary of the bag.

As it is well known, in the context of cosmology, it is very difficult to understand the smallness of the observed present vacuum energy density.
This "cosmological constant problem", has been reformulated in the context of the two measures theory (TMT) \cite{TMT1}, \cite{TMT2}, \cite{TMT3}, \cite{TMT4}, \cite{TMT5}
and more specifically in the context of the scale invariant
realization of such theories  \cite{TMT2}, \cite{TMT3}, \cite{TMT4}, \cite{TMT5}. These theories
can provide a new approach to  the cosmological constant problem and can be generalized to obtain also
a theory with a dynamical spacetime \cite{dyn} . The TMT models consider two measures of integration in the action,
the standard $\sqrt{-g}$ where $g$ is the determinant of the metric and anothe measure $\Phi$ independent
of the metric. To implement scale invariance (S.I.), a dilaton  field is introduced \cite{TMT2}, \cite{TMT3}, \cite{TMT4}, \cite{TMT5}.
  
In the TMT theories we obtain drastic modifications of the dynamics of vaccum energy density, which produces naturally a 
zero cosmological constant and together with this regions of very small vacuum energy density. These ideas work particularly well in the context of scale invariance which can be spontaneously broken by the integration of the equations of motion. What is most important for the present research is that it is the nature of the two measures theories to change not only the dynamics of the vacuum energy density, but also that of the matter itself. For example, in the context of the spontaneouly broken scale invariant theories, the dilaton field decouples from the fermionic matter at high densities, avoiding the fifth forth problem, see first and third references of \cite{TMT4}. On the opposite limit, fermionic matter was shown to contribute to the dark energy density for very small densities, see last reference of \cite{TMT4}.

In this paper our focus will be on the interplay between gauge field dynamics, in particular confinement properties, and vacuum energy density in the context of TMT and whether the intriging questions posed by the MIT bag model can be addressed in this context, in particular why an unconfined phase is associated with a high value of the vacuum energy density, while the unconfined phase appears in a lower energy density state. Interestingly enough, the answer is yes, the TMT model can provide naturally with the basic qualitative features, but more specfic details work out a little different than in the MIT bag model.

 Using the first order formalism, curvature ( $\Phi R$ and  $\sqrt{-g}R^{2}$ ) terms , gauge field terms and dilaton kinetic 
  terms can be introduced in a conformally invariant way. Exponential potentials for the dilaton break down (softly) the conformal invariance down to global scale invariance, which also suffers s.s.b. after integrating the equations of motion. As a result of the s.s.b. of scale invariance phases with different vacuum energy density appear. 
In this paper we will review the principles
of the TMT and in particular the model studied in \cite{TMT2}, which has
global scale invariance. Then, we
look at the generalization of this model \cite{TMT5} by adding a curvature square or simply "$R^{2}$ term"
and show that the resulting model contains now two flat
regions. The existence of two flat regions for the potential
is shown to be consequence of the s.s.b. of the scale symmetry. The model is then further extended to include gauge fields. A gauge field strength squared term coupled to $\sqrt{-g}$,  a square root of a gauge field strength squared term coupled to $\Phi$ and a mass term for the gauge fields coupled to $\Phi$ are the unique candidates which respect local conformal invariance and they can provide a consistent framework to answer the questions posed. For the issue of electric confinement we disregard the mass term and consider only the gauge field strength squared term coupled to $\sqrt{-g}$ and  the square root of a gauge field strength squared term coupled to $\Phi$. This square root term has been studied before in order to reproduce confinement behavior 
\cite{Guendelman},\cite{GGS},
\cite{Guendelman-Gaete},
\cite{interplay},
\cite{Korover}. In the context of the softly broken conformally invariant TMT model it appears however in a particularly natural way. After s.s.b. of scale invariance, the amazing feature that the square root gauge field term is totally screened in the high vacuum energy regions (inside the bags) and acts only outside the bags, reproducing basic qualitative behavior postulated in the M.I.T bag model, also some difficulties present in the original formulation of the square root gauge fields approach to confinement are resolved when the square root term is embedded in the TMT model presented here.

\section{The principles of the Two Measures Theories}

The general structure of general coordinate invariant theories is taken usually as
\begin{equation}\label{e1}
S_{1} = \int{L_{1}}\sqrt{-g} d^{4}x 
\end{equation}
where $g =  det (g_{\mu\nu})$.  The introduction of $\sqrt{-g}$ is required since $d^{4}x$ by itself is not a scalar but the product
$\sqrt{-g} d^{4} x$ is a scalar. Inserting $\sqrt{-g}$,
which has the transformation properties of a
density, produces a scalar action $S_{1}$, as defined by eq.\ref{e1}, provided $L_{1}$ is a scalar.

    In principle nothing prevents us from considering other densities instead of
$\sqrt{-g}$. One costruction of such alternative "measure of integration", is obtained as follows:
 given 4-scalars $\varphi_{a}$ (a =
1,2,3,4), one can construct the density
\begin{equation}\label{e2}
\Phi =  \varepsilon^{\mu\nu\alpha\beta}  \varepsilon_{abcd}
\partial_{\mu} \varphi_{a} \partial_{\nu} \varphi_{b} \partial_{\alpha}
\varphi_{c} \partial_{\beta} \varphi_{d}
\end{equation}
and consider in addition to the action $S_{1}$, as defined by eq.\ref{e1},$S_{2}$, defined as
\begin{equation}\label{e3}
S_{2} =  \int L_{2} \Phi d^{4} x
\end{equation}
$L_{2}$ is again some scalar, which may contain the curvature (i.e. the
gravitational contribution) and a matter contribution, as it can be the case for $S_{1}$, as defined by eq.(\ref{e1}).

    In the action $S_{2}$ defined by eq.\ref{e3} the measure carries degrees of freedom
independent of that of the metric and that of the matter fields. The most
natural and successful formulation of the theory is achieved when the
connection is also treated as an independent degree of freedom. This is
what is usually referred to as the first order formalism.

    One can notice that $\Phi$ is the total derivative of something,
for
example, one can write
\begin{equation}\label{e4}
\Phi = \partial_{\mu} ( \varepsilon^{\mu\nu\alpha\beta}
\varepsilon_{abcd} \varphi_{a}
  \partial_{\nu} \varphi_{b}
\partial_{\alpha}
\varphi_{c} \partial_{\beta} \varphi_{d}).
\end{equation}

    This means that a shift of the form
\begin{equation}\label{e5}
        L_{2} \rightarrow  L_{2}  +  constant
\end{equation}
just adds the integral of a total divergence to the action (\ref{e3}) and it does
not affect therefore the equations of motion of the theory. The same
shift, acting on (\ref{e1}) produces an additional term which gives rise to a
cosmological constant.

    One can consider both contributions, and allowing therefore both geometrical
objects to enter the theory and take as our action
\begin{equation}\label{e6}
S = \int L_{1} \sqrt{-g}d^{4}x + \int L_{2} \Phi  d^{4} x    
\end{equation}

 Here $L_{1}$ and
$L_{2}$ are
$\varphi_{a}$  independent.
There is a good reason not to consider non linear terms in  $\Phi$ that mix $\Phi$ with $\sqrt{-g}$ like
for example
\begin{equation}\label{e7}
\frac{\Phi^{2}}{\sqrt{-g}}
\end{equation}
appear.

    This is because $S$ in eq.(\ref{e6}) is invariant (up to the integral of a total
divergence) under the infinite dimensional symmetry
\begin{equation}\label{e8}
\varphi_{a} \rightarrow \varphi_{a}  +  f_{a} (L_{2})
\end{equation}
where $f_{a} (L_{2})$ is an arbitrary function of $L_{2}$ if $L_{1}$ and
$L_{2}$ are $\varphi_{a}$
independent. Such symmetry (up to the integral of a total divergence) is
absent if mixed terms (like (\ref{e7})) are present.  Therefore (\ref{e6}) is considered
for the case when no dependence on the measure fields (MF) appears in
$L_{1}$ or $L_{2}$.

\section{Conformal invariance softly broken to global scale invariance}
    We will study now the dynamics of a scalar field $\phi$ interacting
with gravity as given by the following action, where except for the potential terms
$U$ and $V$ we have conformal invariance, the potential terms
$U$ and $V$ break down this to global scale invariance.

\begin{equation}\label{e9}
S_{L} =    \int L_{1} \sqrt{-g}   d^{4} x +  \int L_{2} \Phi d^{4} x 
\end{equation}
\begin{equation}\label{e10}
L_{1} = U(\phi)
\end{equation}

\begin{equation}\label{e11}
L_{2} = \frac{-1}{\kappa} R(\Gamma, g) + \frac{1}{2} g^{\mu\nu}
\partial_{\mu} \phi \partial_{\nu} \phi - V(\phi)
\end{equation}
\begin{equation}\label{e12}
R(\Gamma,g) =  g^{\mu\nu}  R_{\mu\nu} (\Gamma) , R_{\mu\nu}
(\Gamma) = R^{\lambda}_{\mu\nu\lambda}
\end{equation}
\begin{equation}\label{e13}
R^{\lambda}_{\mu\nu\sigma} (\Gamma) = \Gamma^{\lambda}_
{\mu\nu,\sigma} - \Gamma^{\lambda}_{\mu\sigma,\nu} +
\Gamma^{\lambda}_{\alpha\sigma}  \Gamma^{\alpha}_{\mu\nu} -
\Gamma^{\lambda}_{\alpha\nu} \Gamma^{\alpha}_{\mu\sigma}.
\end{equation}

The suffix $L$ in $S_{L}$ is to emphasize that here the curvature appears
only linearly. Here,  except for the potential terms
$U$ and $V$ we have conformal invariance, the potential terms
$U$ and $V$ break down this to global scale invariance. Since the breaking of local 
conformal invariance is only through potential terms, we call this a "soft breaking".

    In the variational principle $\Gamma^{\lambda}_{\mu\nu},
g_{\mu\nu}$, the measure fields scalars
$\varphi_{a}$ and the "matter" - scalar field $\phi$ are all to be treated
as independent
variables although the variational principle may result in equations that
allow us to solve some of these variables in terms of others.

For the case the potential terms
$U=V=0$ we have local conformal invariance

\begin{equation}\label{e14}
g_{\mu\nu}  \rightarrow   \Omega(x)  g_{\mu\nu}
\end{equation}

and $\varphi_{a}$ is transformed according to
\begin{equation}\label{e15}
\varphi_{a}   \rightarrow   \varphi^{\prime}_{a} = \varphi^{\prime}_{a}(\varphi_{b})
\end{equation}

\begin{equation}\label{e16}
\Phi \rightarrow \Phi^{\prime} = J(x) \Phi    
\end{equation}
 where $J(x)$  is the Jacobian of the transformation of the $\varphi_{a}$ fields.

This will be a symmetry in the case $U=V=0$ if 
\begin{equation}\label{e17}
\Omega = J
\end{equation}
Notice that $J$ can be a local function of space time, this can be arranged by performing for the 
$\varphi_{a}$ fields one of the (infinite) possible diffeomorphims in the internal $\varphi_{a}$ space.

 We can "softly break" (that is break through potential terms, not kinetic terms) the conformal invariance but still retain a  global
scale invariance in model for very special exponential form for the $U$ and $V$ potentials. Indeed, if we perform the global
scale transformation ($\theta$ =
constant)
\begin{equation}\label{e18}
g_{\mu\nu}  \rightarrow   e^{\theta}  g_{\mu\nu}
\end{equation}
then \ref{e9} is invariant provided  $V(\phi)$ and $U(\phi)$ are of the
form  \cite{TMT2}
\begin{equation}\label{e19}
V(\phi) = f_{1}  e^{\alpha\phi},  U(\phi) =  f_{2}
e^{2\alpha\phi}
\end{equation}
and $\varphi_{a}$ is transformed according to
\begin{equation}\label{e20}
\varphi_{a}   \rightarrow   \lambda_{ab} \varphi_{b}
\end{equation}
which means
\begin{equation}\label{e21}
\Phi \rightarrow det( \lambda_{ab}) \Phi \\ \equiv \lambda
\Phi     \end{equation}
such that
\begin{equation}\label{e22}
\lambda = e^{\theta}
\end{equation}
and
\begin{equation}\label{e23}
\phi \rightarrow \phi - \frac{\theta}{\alpha}.
\end{equation}

\section{Spontaneously broken scale invariance}
We will now work out the equations of motion after introducing $V(\phi)$ and $U(\phi)$
and see how the integration of the equations of motion allows the spontaneous breaking of the
scale invariance.  

    Let us begin by considering the equations which are obtained from
the variation of the fields that appear in the measure, i.e. the
$\varphi_{a}$
fields. We obtain then
\begin{equation}\label{e24}
A^{\mu}_{a} \partial_{\mu} L_{2} = 0
\end{equation}
where  $A^{\mu}_{a} = \varepsilon^{\mu\nu\alpha\beta}
\varepsilon_{abcd} \partial_{\nu} \varphi_{b} \partial_{\alpha}
\varphi_{c} \partial_{\beta} \varphi_{d}$. Since it is easy to
check that  $A^{\mu}_{a} \partial_{\mu} \varphi_{a^{\prime}} =
\frac{\delta aa^{\prime}}{4} \Phi$, it follows that
det $(A^{\mu}_{a}) =\frac{4^{-4}}{4!} \Phi^{3} \neq 0$ if $\Phi\neq 0$.
Therefore if $\Phi\neq 0$ we obtain that $\partial_{\mu} L_{2} = 0$,
 or that
\begin{equation}\label{e25}
L_{2} = \frac{-1}{\kappa} R(\Gamma,g) + \frac{1}{2} g^{\mu\nu}
\partial_{\mu} \phi \partial_{\nu} \phi - V = M
\end{equation}
where M is constant. Notice that this equation breaks spontaneously the global scale invariance of the theory, 
since the left hand side has a non trivial transformation under the scale transformations, while the right
hand side is equal to $M$, a constant that after we integrate the equations is fixed, cannot be changed and therefore
for any $M\neq 0$ we have obtained indeed, spontaneous breaking of scale invariance.

\section{The Equations of Motion in The Einstein Frame}
We will see what is the connection now. As we will see, the connection appears in the original frame as a
non Riemannian object. However, we will see that by a simple conformal tranformation of the metric we can recover
the Riemannian structure. The interpretation of the equations in the frame gives then an interesting physical picture, 
as we will see.

    Let us begin by studying the equations obtained from the variation of the
connections $\Gamma^{\lambda}_{\mu\nu}$.  We obtain then
\begin{equation}\label{e26}
-\Gamma^{\lambda}_{\mu\nu} -\Gamma^{\alpha}_{\beta\mu}
g^{\beta\lambda} g_{\alpha\nu}  + \delta^{\lambda}_{\nu}
\Gamma^{\alpha}_{\mu\alpha} + \delta^{\lambda}_{\mu}
g^{\alpha\beta} \Gamma^{\gamma}_{\alpha\beta}
g_{\gamma\nu}\\ - g_{\alpha\nu} \partial_{\mu} g^{\alpha\lambda}
+ \delta^{\lambda}_{\mu} g_{\alpha\nu} \partial_{\beta}
g^{\alpha\beta}
 - \delta^{\lambda}_{\nu} \frac{\Phi,_\mu}{\Phi}
+ \delta^{\lambda}_{\mu} \frac{\Phi,_           \nu}{\Phi} =  0
\end{equation}
If we define $\Sigma^{\lambda}_{\mu\nu}$    as
$\Sigma^{\lambda}_{\mu\nu} =
\Gamma^{\lambda}_{\mu\nu} -\{^{\lambda}_{\mu\nu}\}$
where $\{^{\lambda}_{\mu\nu}\}$   is the Christoffel symbol, we
obtain for $\Sigma^{\lambda}_{\mu\nu}$ the equation
\begin{equation}\label{e27}
    -  \sigma, _{\lambda} g_{\mu\nu} + \sigma, _{\mu}
g_{\nu\lambda} - g_{\nu\alpha} \Sigma^{\alpha}_{\lambda\mu}
-g_{\mu\alpha} \Sigma^{\alpha}_{\nu \lambda}
+ g_{\mu\nu} \Sigma^{\alpha}_{\lambda\alpha} +
g_{\nu\lambda} g_{\alpha\mu} g^{\beta\gamma} \Sigma^{\alpha}_{\beta\gamma}
= 0
\end{equation}
where  $\sigma = ln \chi, \chi = \frac{\Phi}{\sqrt{-g}}$.

    The general solution of eq.(\ref{e28}) is
\begin{equation}\label{e28}
\Sigma^{\alpha}_{\mu\nu} = \delta^{\alpha}_{\mu}
\lambda,_{\nu} + \frac{1}{2} (\sigma,_{\mu} \delta^{\alpha}_{\nu} -
\sigma,_{\beta} g_{\mu\nu} g^{\alpha\beta})
\end{equation}
where $\lambda$ is an arbitrary function due to the $\lambda$ - symmetry
of the
curvature \cite{Lambda} $R^{\lambda}_{\mu\nu\alpha} (\Gamma)$,
\begin{equation}\label{e29}
\Gamma^{\alpha}_{\mu\nu} \rightarrow \Gamma^{\prime \alpha}_{\mu\nu}
 = \Gamma^{\alpha}_{\mu\nu} + \delta^{\alpha}_{\mu}
Z,_{\nu}
\end{equation}
Z  being any scalar (which means $\lambda \rightarrow \lambda + Z$).

    If we choose the gauge $\lambda = \frac{\sigma}{2}$, we obtain
\begin{equation}\label{e30}
\Sigma^{\alpha}_{\mu\nu} (\sigma) = \frac{1}{2} (\delta^{\alpha}_{\mu}
\sigma,_{\nu} +
 \delta^{\alpha}_{\nu} \sigma,_{\mu} - \sigma,_{\beta}
g_{\mu\nu} g^{\alpha\beta}).
\end{equation}

    Considering now the variation with respect to $g^{\mu\nu}$, we
obtain
\begin{equation}\label{e31}
\Phi (\frac{-1}{\kappa} R_{\mu\nu} (\Gamma) + \frac{1}{2} \phi,_{\mu}
\phi,_{\nu}) - \frac{1}{2} \sqrt{-g} U(\phi) g_{\mu\nu} = 0
\end{equation}
solving for $R = g^{\mu\nu} R_{\mu\nu} (\Gamma)$  from eq.(\ref{e31}) and introducing
in eq.(\ref{e25}), we obtain
\begin{equation}\label{e32}
M + V(\phi) - \frac{2U(\phi)}{\chi} = 0
\end{equation}
a constraint that allows us to solve for $\chi$,
\begin{equation}\label{e33}
\chi = \frac{2U(\phi)}{M+V(\phi)}.
\end{equation}

    To get the physical content of the theory, it is best
consider variables that have well defined dynamical interpretation. The original
metric does not has a non zero canonical  momenta. The fundamental
variable of the theory in the first order formalism is the connection and its
canonical momenta is a function of $\overline{g}_{\mu\nu}$, given by,

\begin{equation}\label{e34}
\overline{g}_{\mu\nu} = \chi g_{\mu\nu}
\end{equation}

and $\chi$  given by eq.(\ref{e33}). Interestingly enough, working with $\overline{g}_{\mu\nu}$
is the same as going to the "Einstein Conformal Frame".
In terms of $\overline{g}_{\mu\nu}$   the non
Riemannian contribution $\Sigma^{\alpha}_{\mu\nu}$
dissappears from the equations. This is because the connection
can be written as the Christoffel symbol of the metric
$\overline{g}_{\mu\nu}$ .
In terms of $\overline{g}_{\mu\nu}$ the equations
of motion for the metric can be written then in the Einstein
form (we define $\overline{R}_{\mu\nu} (\overline{g}_{\alpha\beta}) =$
 usual Ricci tensor in terms of the bar metric $= R_{\mu\nu}$ and
 $\overline{R}  = \overline{g}^{\mu \nu}  \overline{R}_{\mu\nu}$ )
\begin{equation}\label{e35}
\overline{R}_{\mu\nu} (\overline{g}_{\alpha\beta}) - \frac{1}{2}
\overline{g}_{\mu\nu}
\overline{R}(\overline{g}_{\alpha\beta}) = \frac{\kappa}{2} T^{eff}_{\mu\nu}
(\phi)
\end{equation}
where
\begin{equation}\label{e36}
T^{eff}_{\mu\nu} (\phi) = \phi_{,\mu} \phi_{,\nu} - \frac{1}{2} \overline
{g}_{\mu\nu} \phi_{,\alpha} \phi_{,\beta} \overline{g}^{\alpha\beta}
+ \overline{g}_{\mu\nu} V_{eff} (\phi)
\end{equation}

and
\begin{equation}\label{e37}
V_{eff} (\phi) = \frac{1}{4U(\phi)}  (V+M)^{2}.
\end{equation}

    In terms of the metric $\overline{g}^{\alpha\beta}$ , the equation
of motion of the Scalar
field $\phi$ takes the standard General - Relativity form
\begin{equation}\label{e38}
\frac{1}{\sqrt{-\overline{g}}} \partial_{\mu} (\overline{g}^{\mu\nu}
\sqrt{-\overline{g}} \partial_{\nu}
\phi) + V^{\prime}_{eff} (\phi) = 0.
\end{equation}

    Notice that if  $V + M = 0,  V_{eff}  = 0$ and $V^{\prime}_{eff}
= 0$ also, provided $V^{\prime}$ is finite and $U \neq 0$ there.
This means the zero cosmological constant
state
is achieved without any sort of fine tuning. That is, independently
of whether we add to $V$ a constant piece, or whether we change
the value of $M$, as long as there is still a point
where $V+M =0$, then still $ V_{eff}  = 0$ and $V^{\prime}_{eff} = 0$
( still provided $V^{\prime}$ is finite and $U \neq 0$ there).
This is the basic feature
that characterizes the TMT and allows it to solve the 'old'
cosmological constant problem, at least at the classical level.

    In what follows we will study the effective potential (\ref{e37}) for the special case of global
scale invariance, which as we will see displays additional very special
features which makes it attractive in the context of cosmology.

    Notice that in terms of the variables $\phi$,
$\overline{g}_{\mu\nu}$, the "scale"
transformation becomes only a shift in the scalar field $\phi$, since
$\overline{g}_{\mu\nu}$ is
invariant (since $\chi \rightarrow \lambda^{-1} \chi$  and $g_{\mu\nu}
\rightarrow \lambda g_{\mu\nu}$)
\begin{equation}\label{e39}
\overline{g}_{\mu\nu} \rightarrow \overline{g}_{\mu\nu}, \phi \rightarrow
\phi - \frac{\theta}{\alpha}.
\end{equation}

    If $V(\phi) = f_{1} e^{\alpha\phi}$  and  $U(\phi) = f_{2}
e^{2\alpha\phi}$ as
required by scale
invariance eqs.(\ref{e18}),(\ref{e20}),(\ref{e21}),(\ref{e22}),(\ref{e23}), we obtain from the expression (\ref{e37})
\begin{equation}\label{e40}
    V_{eff}  = \frac{1}{4f_{2}}  (f_{1}  +  M e^{-\alpha\phi})^{2}
\end{equation}

    Since we can always perform the transformation $\phi \rightarrow
- \phi$ we can
choose by convention $\alpha > 0$. We then see that as $\phi \rightarrow
\infty, V_{eff} \rightarrow \frac{f_{1}^{2}}{4f_{2}} =$ const.
providing infinite flat region. Also a minimum is achieved at zero
cosmological constant for the case $\frac{f_{1}}{M} < 0$ at the point
\begin{equation}\label{e41}
\phi_{min}  =  \frac{-1}{\alpha} ln \mid\frac{f_1}{M}\mid.
\end{equation}

    Finally, the second derivative of the potential  $V_{eff}$  at the
minimum is
\begin{equation}\label{e42}
V^{\prime\prime}_{eff} = \frac{\alpha^2}{2f_2} \mid{f_1}\mid^{2} > 0
\end{equation}
if
$f_{2} > 0$,
there are many interesting issues that one can raise here. The first one
is of course the fact that a realistic scalar field potential, with
massive exitations when considering the true vacuum state, is achieved in
a way consistent with the idea  of scale
invariance.

    A peculiar feature of the potential (\ref{e40}), is that the absolute
value of the constant $M$, does not affect the physics of the problem, only the
sign will have an effect. This is because if we perform a shift
\begin{equation}\label{e43}
\phi \rightarrow \phi + \Delta
\end{equation}
in the potential (\ref{e40}), this is equivalent to the change in the integration
constant  M
\begin{equation}\label{e44}
M \rightarrow M e^{-\alpha\Delta}.
\end{equation}

    We see therefore that if we change  $M$ in any way, without changing
the sign of M, the only effect this has is to shift the whole potential.
The physics of the potential remains unchanged, however. This is
reminiscent of the dilatation invariance of the theory, which involves
only a shift in $\phi$  if $\overline{g}_{\mu\nu}$   is used (see eq. (\ref{e39})).

    This is very different from the situation for two generic
functions
$U(\phi)$ and
$V(\phi)$ in (\ref{e37}). There, $M$ appears in $V_{eff}$ as a true new parameter
that
generically changes the shape of the potential $V_{eff}$, i.e. it is
impossible
then to compensate the effect of M with just a shift. For example  M will
appear in the value of the second derivative of the potential at the
minimum, unlike what we see in eq. (\ref{e42}), where we see that
$V^{\prime\prime}_{eff}$ (min) is $M$
independent.

    In conclusion, the scale invariance of the original theory is
responsible for the non appearance (in the physics) of a certain scale,
that associated to M. However, masses do appear, since the coupling to two
different measures of $L_{1}$ and $L_{2}$ allow us to introduce two
independent
couplings  $f_{1}$ and $f_{2}$, a situation which is  unlike the
standard
formulation of globally scale invariant theories, where usually no stable
vacuum state exists.

We can compare the appearance of the potential $V_{eff} (\phi)$, which has
privileged
some point depending on M (for example the minimum of the potential will
have to be at some specific point), although the theory has the
"translation invariance" (\ref{e43}), to the physics of solitons.

    In fact, this very much resembles the appearance of solitons in a
space-translation invariant theory: The soliton solution has to be
centered at some point, which of course is not determined by the theory.
The soliton of  course breaks the space translation invariance
spontaneously, just as the existence of the non trivial potential $V_{eff}
(\phi)$
breaks here spontaneously the translations in $\phi$ space, since $V_{eff}
(\phi)$ is
not a constant.

The constant of integration $M$ plays a very important role indeed:
any non vanishing value for this constant implements, already at the
classical level S.S.B. of scale invariance.

\section{ Generation of two flat regions after the introduction of a $R^{2}$ term}

As we have seen, it is possible to obtain a model that through a spontaneous breaking of scale invariace
can give us a flat region. We want to obtain now two flat regions in our effective potential.
A simple generalization of the action $S_{L}$ will fix this.
What one needs to do is simply consider  the
addition of a scale invariant term of the form

\begin{equation}\label{e45}
S_{R^{2}} = \epsilon  \int (g^{\mu\nu} R_{\mu\nu} (\Gamma))^{2} \sqrt{-g} d^{4} x
\end{equation}

The total action being then $S = S_{L} + S_{R^{2}}$.
In the first order formalism $ S_{R^{2}}$ is not only globally scale invariant
but also locally scale invariant, that is conformally invariant (recall that
in the first order formalism the connection is an independent degree of freedom
and it does not transform under a conformal transformation of the metric). Many features are different 
when comparing with non linear in the curvature theories formulated in the second order formalism \cite{barrow},
\cite{mijic}, in particular the order of the equations.

Let us see what the equations of motion tell us, now with the addition of
$S_{R^{2}}$ to the action. First of all, since the addition has been only to
the part of the action that couples to $ \sqrt{-g}$, the equations of motion
derived from the variation of the measure fields remains unchanged. That is
eq.(\ref{e25}) remains valid.

The variation of the action with respect to $ g^{\mu \nu}$ gives now

\begin{equation}\label{e46}
 R_{\mu\nu} (\Gamma) ( \frac{-\Phi}{\kappa} + 2 \epsilon R  \sqrt{-g}) +
\Phi \frac{1}{2} \phi,_{\mu} \phi,_{\nu} -
\frac{1}{2}(\epsilon R^{2} + U(\phi)) \sqrt{-g} g_{\mu\nu} = 0
\end{equation}

It is interesting to notice that if we contract this equation with
 $ g^{\mu \nu}$ , the $\epsilon$ terms do not contribute. This means
that the same value for the scalar curvature $R$ is obtained as in section 2,
 if we express our result in terms of $\phi$, its derivatives and
$ g^{\mu \nu}$ .
Solving the scalar curvature from this and inserting in the other
$\epsilon$ - independent equation $L_{2} = M$  we get still the same
solution for the ratio of the measures which was found in the case where
the $\epsilon$ terms were absent,
i.e. $\chi =  \frac{\Phi}{\sqrt{-g}}  = \frac{2U(\phi)}{M+V(\phi)}$.

In the presence of the $\epsilon R^{2} $ term in the action, eq. (\ref{e26})
gets modified so that instead of $\Phi$,  $\Omega$  =
$\Phi - 2 \epsilon R \sqrt{-g}$ appears. This in turn implies that
eq.(\ref{e27}) mantains its form but where $\sigma$ is replaced by
$\omega  = ln (\frac{\Omega}{\sqrt{-g}}) =
 ln ( \chi -2\kappa \epsilon R)$,
where once again,
$\chi =  \frac{\Phi}{\sqrt{-g}} = \frac{2U(\phi)}{M+V(\phi)}$.

Following then the same steps as in the model without the curvature square terms, we can then verify that the
connection is the Christoffel symbol of the metric $\overline{g}_{\mu\nu}$
given by

\begin{equation}\label{e47}
\overline{g}_{\mu\nu}   = (\frac{\Omega}{\sqrt{-g}}) g_{\mu\nu}
 = (\chi -2\kappa \epsilon R) g_{\mu\nu}
\end{equation}

$\overline{g}_{\mu\nu} $ defines now the "Einstein frame". Equations (\ref{e46})
can now be expressed in the "Einstein form"

\begin{equation}\label{e48}
\overline{R}_{\mu\nu} -  \frac{1}{2}\overline{g}_{\mu\ \nu} \overline{R} =
\frac{\kappa}{2} T^{eff}_{\mu\nu}
\end{equation}

where

\begin{equation}
 T^{eff}_{\mu\nu} =\label{e49}
\frac{\chi}{\chi -2 \kappa \epsilon R} (\phi_{,\mu} \phi_{,\nu} - \frac{1}{2} \overline
{g}_{\mu\nu} \phi_{,\alpha} \phi_{,\beta} \overline{g}^{\alpha\beta})
+ \overline{g}_{\mu\nu} V_{eff}
\end{equation}

where

\begin{equation}\label{e50}
 V_{eff}  = \frac{\epsilon R^{2} + U}{(\chi -2 \kappa \epsilon R)^{2} }
\end{equation}

Here it is satisfied that $\frac{-1}{\kappa} R(\Gamma,g) +
\frac{1}{2} g^{\mu\nu}\partial_{\mu} \phi \partial_{\nu} \phi - V = M $,
equation that expressed in terms of $ \overline{g}^{\alpha\beta}$
 becomes

$\frac{-1}{\kappa} R(\Gamma,g) + (\chi -2\kappa \epsilon R)
\frac{1}{2} \overline{g}^{\mu\nu}\partial_{\mu} \phi \partial_{\nu} \phi - V = M$.
 This allows us to solve for $R$ and we get,

\begin{equation}\label{e51}
R = \frac{-\kappa (V+M) +\frac{\kappa}{2} \overline{g}^{\mu\nu}\partial_{\mu} \phi \partial_{\nu} \phi \chi}
{1 + \kappa ^{2} \epsilon \overline{g}^{\mu\nu}\partial_{\mu} \phi \partial_{\nu} \phi}
\end{equation}

Notice that
 if we express $R$ in
terms of $\phi$, its derivatives and $ g^{\mu \nu}$, the result is the
same as in the model without the curvature squared term, this is not true anymore once we express
 $R$ in terms of $\phi$, its derivatives and $\overline{g}^{\mu\nu}$.

In any case, once we insert (\ref{e51}) into (\ref{e50}),
we see that the  effective potential  (\ref{e50}) will depend on the derivatives of the
scalar field now. It acts as a normal scalar field potential under the
conditions of slow rolling  or low gradients and in the case the
scalar field is near the region $M+V(\phi ) = 0$.

Notice that since
$\chi =   \frac{2U(\phi )}{M+V(\phi )}$,
then if ${M+V(\phi) = 0}$, then, as in the simpler model without the curvature squared terms, we obtain that
 $ V_{eff}  =  V'_{eff} =  0$ at that point without fine tuning
(here by $ V'_{eff}$ we mean the derivative  of $ V_{eff}$ with
respect to the scalar field $\phi$, as usual).

In the case of the scale invariant case, where $V$ and $U$ are given by
equation (\ref{e19}), it is interesting to study the shape of $ V_{eff} $
as a function of $\phi$
in the case of a constant $\phi$, in which case $ V_{eff} $ can be
regarded as a real scalar field potential. Then from (\ref{e51}) we get
$R = -\kappa (V+M)$, which inserted in (\ref{e50}) gives,
\begin{equation}\label{effpotslow}
 V_{eff}  =
\frac{(f_{1} e^{ \alpha \phi }  +  M )^{2}}{4(\epsilon \kappa ^{2}(f_{1}e^{\alpha \phi}  +  M )^{2} + f_{2}e^{2 \alpha \phi })}
\end{equation}

The limiting values of $ V_{eff} $ are:

First, for asymptotically
large positive values, we have,
\begin{equation}\label{phipos}
V_{eff}( \phi \rightarrow  \infty ) \rightarrow
\frac{f_{1}^{2}}{4(\epsilon \kappa ^{2} f_{1}^{2} + f_{2})} 
\end{equation}

Second, for asymptotically large but negative values of the scalar field, we have,
\begin{equation}\label{phineg}
 V_{eff} (\phi \rightarrow - \infty )\rightarrow \frac{1}{4\epsilon \kappa ^{2}}
\end{equation}

In these two asymptotic regions ($\alpha \phi \rightarrow  \infty  $
and $\alpha \phi \rightarrow - \infty  $) an examination of the scalar
field equation reveals that a constant scalar field configuration is a
solution of the equations, as is of course expected from the flatness of
the effective potential in these regions.

Notice that in all the above discussion it is fundamental that $ M\neq 0$.
If $M = 0$ the potential becomes just a flat one,
$V_{eff} = \frac{f_{1}^{2}}{4(\epsilon \kappa ^{2} f_{1}^{2} + f_{2})}$
everywhere (not only at high values  of $\alpha \phi$). All the non trivial
features, the other flat
region and the minimum at zero if $M<0$ are all lost .
As we discussed in the model without a curvature squared term, $ M\neq 0$ implies the we are considering a
situation with S.S.B. of scale invariance.

For  $M>0$ , the absolute minimum of the effective potential is found for $\phi \rightarrow  \infty $.

\section{ The Bag Constant}
In this short section, even before we add gauge fields into our model we want to define the regions which we will call "inside the bag" and the region we will call "outside the bag". 

Here we will use the most straightforward approach 
and take  the coupling constants $\epsilon$ and  $f_{2}$ to be positive, then we see that automatically, the energy
density obtained for  high values  of $\alpha \phi$, $\frac{f_{1}^{2}}{4(\epsilon \kappa ^{2} f_{1}^{2} + f_{2})}$  is smaller than the value of the energy density obtained for extremelly negative values of  $\alpha \phi$, $ \frac{1}{4\epsilon \kappa ^{2}}$ .  This is so since,

\begin{equation}\label{inequality}
0< \frac{f_{1}^{2}}{4(\epsilon \kappa ^{2} f_{1}^{2} + f_{2})} =\\
\frac{1}{4(\epsilon \kappa ^{2}  + f_{2}/f_{1}^{2})} < \frac{1}{4\epsilon \kappa ^{2}}
\end{equation}

The value $ \frac{1}{4\epsilon \kappa ^{2}}$
will define the energy of the false vacuum, or the "bag constant". Stricktly speaking, the "bag constant" should be the difference of the vacuum energy inside $ \frac{1}{4\epsilon \kappa ^{2}}$
 minus the vacuum energy outside $\frac{f_{1}^{2}}{4(\epsilon \kappa ^{2} f_{1}^{2} + f_{2})}$, but we will consider this vacuum energy density outside completly negligible compare to the vacuum energy inside $ \frac{1}{4\epsilon \kappa ^{2}}$.

From \ref{inequality}, we see that the asymmetry between the two vacuum energies depend of how big is $f_{2}/f_{1}^{2}$ compared to 
$4\epsilon \kappa ^{2}$.

Notice that we are allowed to take the limit $\kappa \rightarrow 0$ and still retail a meaningfull bag constant, if the bag constant is kept  fixed, which implies that 
$\epsilon \kappa ^{2}$ is kept fixed.

We will be concerned also in the first place with the case in which $f_{1}$ and $M$ are also taken to be positive, for reasons  to be studied in the following sections.

\section{Effective Action in the Einstein Frame}

We start considering the case where $\phi_{,\alpha}$ is a time like vector, but then drop this condition. If $\phi_{,\alpha}$ is a time like vector, then the effective energy-momentum tensor can be
represented in a form like that of  a perfect fluid in the case,
\begin{equation}
T_{\mu\nu}^{eff}=(\rho +p)u_{\mu}u_{\nu}-p\bar{g}_{\mu\nu},
\qquad \text{where} \qquad
u_{\mu}=\frac{\phi_{,\mu}}{(2X)^{1/2}}\label{Tmnfluid}
\end{equation}
here $X\equiv\frac{1}{2}\bar{g}^{\alpha\beta}\phi_{,\alpha}\phi_{,\beta}$. This defines a pressure functional and an energy density functional.
The system of equations obtained after solving for $\chi$, working 
in the Einstein frame with the metric
$\bar{g}_{\mu \nu}$ can be obtained from a
 "k-essence" type effective action, as it is standard in cosmological treatments 
 of theories with non linear kinetic tems or k-essence models\cite{k-essence}. The action from which the classical equations follow is, 
\begin{equation}
S_{eff}=\int\sqrt{-\bar{g}}d^{4}x\left[-\frac{1}{\kappa}\bar{R}(\bar{g})
+p\left(\phi,R,X\right)\right] \label{k-eff}
\end{equation}

\begin{equation}
 p = \frac{\chi}{\chi -2 \kappa \epsilon R}X - V_{eff}
\end{equation}

\begin{equation}\label{e50N}
 V_{eff}  = \frac{\epsilon R^{2} + U}{(\chi -2 \kappa \epsilon R)^{2} }
\end{equation}

where it is understood that,
\begin{equation}\label{chi2}
\chi = \frac{2U(\phi)}{M+V(\phi)}.
\end{equation}

We can in fact forget now about the condition of $\phi_{,\alpha}$ being a time like vector and see that indeed the above action reproduces the correct equations regards of whether $\phi_{,\alpha}$ is timelike, lightlike or spacelike. So we take  (\ref{k-eff}) as the action for the theory in the Einstein frame without any restriction on the nature of $\phi_{,\alpha}$ .
We have two possible formulations concerning $R$:
Notice first that $\bar{R}$ and $R$ are different objects, the $\bar{R}$ is the Riemannian curvature scalar in the Einstein frame,
while $R$ is a different object. This $R$ will be treated in two different ways:

1. First order formalism for $R$. Here $R$ is a lagrangian variable, determined as follows,  $R$ that appear in the expression above for $p$ can be obtained from the variation of the pressure functional action above with respect to $R$, this gives exactly the expression for $R$ that has been solved already in terms of $X, \phi$, etc. 

2. Second order formalism for $R$. $R$ that appear in the action above is exactly the expression for $R$ that has been solved already in terms of $X, \phi$, etc. The second order formalism can be obtained from the first order formalism by solving algebraically R from the eq. obtained by variation of $R$ , and inserting back into the action.

\section{ "Confinement Terms": basic ideas}

Strong interaction dynamics involves many remarkable phenomena:
first, QCD at very short distances has the asymptotically free
property \cite{free}. In contrast to this, in the infrared region,
we are should obtain the confined phase, where only color singlets
survive, since no free quarks or gluons have been observed so far.
Also at high temperatures, we expect a deconfinement phase
transition \cite{deconf}.

Although the asymptotically free property is clearly understood
theoretically, the  confinement is not. Lattice gauge theory
provides nevertheless numerical evidence for  confinement in the
context of QCD \cite{lattice}.

Given the theoretical difficulties concerning the description of
the confining phase, several phenomenological approaches have been
developed , for example the Cornell potential for  heavy quarks
\cite{Cornell} and the MIT bag model \cite{MIT}, which gives a
comprehensive description of hadron spectroscopy.

An interesting type of field theories can be explored in connection with this problem.
These are theories where a square root of a gauge field strength square is introduced. These
models allow, when no other term is introduced in the gauge field action, to delta function string like solutions \cite{NielSpa}.
Also more regular type of flux tube solutions have been studied in a square root of a gauge field strength square field theory \cite{Amer} .

In Ref. \cite{Guendelman}  a model was proposed where the
spontaneous symmetry breaking (ssb) of scale invariance induces an
effective dynamics including a square root of a gauge field square field strength, which when
combined with a regular gauge field term, gives rise to confinement. Here the ssb
of scale invariance originates from the dynamics of  maximal rank
gauge fields. The resulting theory that results in fact satisfies
the requirements studied by 't Hooft \cite{pert-conf} for
perturbative confinement. In this context, it is an example of a classical model for confinement see for example
\cite{Piran} and \cite{Lehmann} for a general review, although in our case the effective action is highly constrained by symmetry considerations.

More explicit computations in the context of this model
\cite{Guendelman-Gaete} have shown that it gives rise to the
Cornell confining potential for static sources\cite{Cornell}. This
model can be used also to study the interplay between confinement
effects and ssb of gauge symmetry \cite{interplay} . 

In this paper we will see that incorporating the ideas of
\cite{Guendelman} embedding them in the scale invariant framework of TMT theories
allows to enlarge the set of phenomena described. In addition to
confinement we can also obtain bag structures where a deconfined
phase is obtained in a central region of a bag, while confinement is
obtained outside.

We review now the connection between global scale
symmetry breaking and confinement as it was originally introduced in
Ref.\cite{Guendelman}, in the next chapter we will discuss
how all this is reformulated in order to embedded it in the TMT model. 
For this purpose we restrict our attention
to the flat space time action,
\begin{equation}
S_{YM} = \int {d^4 } x\left( - \frac{1}{4}F_{\mu \nu }^{a} F^{a\mu
\nu}
 \right)\ ,\label{YM}
\end{equation}
where $ F_{\mu \nu }^a  = \partial _\mu  A_\nu ^a - \partial _\nu
A_\mu ^a + e f^{abc} A_\mu ^b A_\nu ^c$, where $e$ is the gauge
coupling constant.
 This theory is invariant
under the scale symmetry
\begin{equation}
A^{a}_\mu  \left( x \right) \mapsto A_\mu ^ {a\prime}  \left( x
\right) = \lambda^{-1}\, A^{a}_\mu  \left( \lambda\, x \,\right),
\end{equation}
where $\lambda$ is a constant.

Let us introduce  the following action, which is at this point a new but
equivalent action to the action above
\begin{eqnarray}
{\mathcal{S}_1}\left[\, \omega\ ,  A^{a}_\mu\ \right]
 =&& \int {d^4 } x\left[\,  - \frac{1}{4}\omega ^2
+\frac{1}{2}\omega \sqrt { - F_{\mu \nu }^{a}\, F^{a\mu \nu } }
\right]\,  \label{ma}
\end{eqnarray}

 where, $\omega$  is an
auxiliary scalar field,  its scaling transformation is
\begin{eqnarray}
&&\omega  \mapsto \lambda ^{-2} \omega \left( \lambda\, x
\,\right)\label{cr25c}
\end{eqnarray}

Upon solving the equation of motion associated to the variation
with respect to $\omega$, we obtain that $\omega = \sqrt { -
F_{\mu \nu }^{a}\, F^{a\mu \nu }}$ and inserting back into (\ref{ma}), we
obtain once again (\ref{YM}).

Let us now declare that $\omega$ is not a fundamental field, but
that instead
\begin{equation}
\omega = \epsilon^{\lambda\mu\nu\rho}
\partial_{[\,\lambda}\, A_{\mu\nu\rho \,]}
\end{equation}
Now the variation with respect to $ A_{\mu\nu\rho }$ gives rise to
the equation:
\begin{eqnarray}
 \epsilon^{\lambda\mu\nu\rho} \partial_{\lambda}
 \left[\ \omega - \sqrt { - F_{\mu \nu }^{a}\, F^{a\mu \nu } }\right]\ = 0
\end{eqnarray}

Which is satisfied by,
\begin{eqnarray}
\omega = \sqrt { - F_{\mu \nu }^{a}\, F^{a\mu \nu }} + N
\end{eqnarray}

here $N$ is a space time constant which produces the ssb of scale
invariance and it is in fact associated with the spontaneous
generation of confining behavior. This can be seen by considering
the equations of motion in the case $N\neq 0$

\begin{equation}
 \nabla _ \mu
\left[ {\left( {\sqrt { - F_{\alpha \beta }^a F^{a\alpha \beta } }
+ N} \right)\frac{{F^{a\mu \nu } }}{{\sqrt { - F_{\alpha \beta }^b
F^{b\alpha \beta } } }}} \right] = 0. \label{cr50}
\end{equation}

Then, assuming spherical symmetry and time independence, we obtain
that, in addition to a Coulomb like piece, a linear term
proportional to $N$ is obtained for $A^{a 0}$. A quantum
computation also confirms this result \cite{Guendelman-Gaete}.

Furthermore, these equations are consistent with the 't Hooft
criteria for perturbative confinement. In fact in the infrared
region the above equation implies that

\begin{equation}
{F^{a\mu \nu } = - N \frac{{F^{a\mu \nu } }}{{\sqrt { - F_{\alpha
\beta }^b F^{b\alpha \beta } } }}}. \label{cr51}
\end{equation}

plus negligible terms in the infrared sector. Interestingly
enough, for a static  source, this automatically implies that the
chromoelectric field has a fixed amplitude. Confinement is obvious
then, since in the presence of two external oppositely charged
sources, by symmetry arguments, one can see that such a constant
amplitude chromoelectric field must be in the direction of the
line joining the two charges. The potential that gives rise to
this kind of field configuration is of course a linear potential.
Notice that the above equation implies that $N<0$, otherwise the
electric field would be antiparallel to itself. The string tension
between static quarks of opposite charge is proportional to the
absolute value of $N$.

A configuration with a net charge is seen from the above equation
to produce, at very large distances from the source, an electric
field which is radial but constant in strength. This will in fact
produce an infrared divergent total energy for the system. In
general any non color singlet will be un-physical because these
systems will have infinite energy. Furthermore such configuration
with net charge will detect and be attracted to any opposite
charge no matter how far this will be by a linear potential, so a
system with net charge can never be isolated.

This is the situation for $N<0$, for $N>0$, the contradiction found
by encountering that the field is opposite to itself
means only that this infrared regime is not attainable, at some radious the
solution breaks down \cite{Korover} and a new charge must appear.
In the reformulation of the theory when embedding it in the TMT model 
we find indeed that we must reenter inside into another bag region again. 
So the case $N>0$  provides indeed another picture for cofinement, furthermore,
for this case one can show that free gluons, represented by plane waves, aquire 
infinite energy  and are therefore non existent as physical states\cite{Korover}.

There is an effective action describing the equations with $N \neq 0$, it is obtained by adding a Born-Infeld contribution
to the Yang-Mills action,

\begin{equation}
S_N =
\int d^4x\,\left[\,  - \frac{1}{4} F_{\mu \nu }^{a}\, F^{a\mu \nu }
+\frac{N}{2}\, \sqrt { - F_{\mu \nu }^{a}\, F^{a\mu \nu } }
\,\right],
\end{equation}

in the next section (going back to generally coordinate invariant models) we will see that the incorporation of a term of the form 
$ \sqrt { - F_{\mu \nu }^{a}\, F^{a\mu \nu } }$
in the TMT case is quite natural, in fact, there is a good reason to include it, since it respects conformal symmetry if coupled to the new measure $\Phi$ . This kind of coupling of a square root gauge field strength to a new measure has been considered in the context of conformally invariant braneworld scenarios\cite{sqrt}, which allow compactification, branes and zero four dimensional cosmological constant.
Another place where square root of gauge field square coupled to a  modified measure find a natural place is in the formulation of Weyl invariant brane theories\cite{sqrtbwill}.
\section{ Gauge Field Kinetic Terms, Mass Terms and "Confinement Terms" embedded in the Softly Broken Conformally Invariant TMT Model }
An early model which enriches the "square root" gauge theory with a dilaton field so that it could describe confined and
unconfined regions  (bags) was done "by hand" in \cite{confdeconf}. This will be obtained most elegantly however by embedding the square root terms into the TMT  formalism.

We can see that in the context of the generally coordinate invariant TMT with softly broken conformal invariance a $\sqrt { - F_{\mu \nu }^{a}\, F^{a\mu \nu }}$ is very natural.
The reason for this is very simple: conformally invariant terms (with respect to (\ref{e14}) , (\ref{e15}), (\ref{e16}) and (\ref{e17}))in TMT  are of two kinds, if they multiply the measure $\Phi$ they they must have homogeneity 1 with respect to $g^{\mu \nu}$, or if they multiply the measure $\sqrt{-g}$ they they must have homogeneity 2   with respect to $g^{\mu \nu}$, since 
$\sqrt { - F_{\mu \nu }^{a}\, F^{a\mu \nu }}$ $=\sqrt { - F_{\mu \nu }^{a}\, F^{a}_{\alpha\beta}g^{\mu\alpha}g^{\nu\beta}}$, then according to (\ref{e14}) $\sqrt { - F_{\mu \nu }^{a}\, F^{a\mu \nu }}\rightarrow \Omega^{-1}\sqrt { - F_{\mu \nu }^{a}\, F^{a\mu \nu }}$ if $\Omega>0$
and $\Phi \rightarrow  J \Phi= \Omega \Phi$, so that $\Phi\sqrt { - F_{\mu \nu }^{a}\, F^{a\mu \nu }}$ is invariant, provided $J=\Omega>0$.

A similar story happens with a mass term for the gluon, $A_{\mu }^{a}\, A^{a}_{\alpha}g^{\mu\alpha}$ in TMT, this can be a conformally invariant
if it goes multiplied with the measure $\Phi$.

Likewise, the  conformally invariance implies that a term proportional to
 $F_{\mu \nu }^{a}F^{a}_{\alpha\beta}g^{\mu\alpha}g^{\nu\beta}$ has come multiplied by 
the Riemannian measure $\sqrt{-g}$, since $\sqrt{-g}F_{\mu \nu }^{a}F^{a}_{\alpha\beta}g^{\mu\alpha}g^{\nu\beta}$ is invariant under conformal transformations of the metric. We take therefore for our softly broken conformal invariant model, where we exclude mass terms for the gluons.

\begin{equation}\label{finalaction}
S = S_L + S_{R^2} - \frac  {1}{4}\int d^4x\sqrt{-g}F_{\mu \nu }^{a}F^{a}_{\alpha\beta}g^{\mu\alpha}g^{\nu\beta} 
+\frac{N}{2}\int d^4x \Phi\sqrt{-F_{\mu \nu }^{a}F^{a}_{\alpha\beta}g^{\mu\alpha}g^{\nu\beta}}
\end{equation}
here $S_L$ is defined by equations (\ref{e9}),(\ref{e10}),(\ref{e11}) and $S_{R^2}$ by (\ref{e45}).

In (\ref{finalaction}) the term of the form $\Phi \sqrt { - F_{\mu \nu }^{a}\, F^{a\mu \nu }}$ does not represent a symmetry breaking term, in contrast to the original approach to the square root term, explained in the previous section, where a square root  term appeared as a result of s.s.b. of scale invariance. The introduction of the measure $\Phi$ allows us to consider the square root term while respecting conformal invariance.

\section{Description of the Bag Dynamics in the Softly Broken Conformally Invariant TMT Model }
Let us proceed now to describe the consequences of the action (\ref{finalaction}). 
The steps to follow are the same as in the case where we did not have gauge fields.

One interesting fact is that the terms that enter the constraint that determines $\chi$ are only those that break the
conformal invariance and  the constant of integration $M$.  Since the new terms involving the gauge fields do not break the conformal invariance 
(\ref{e14}), (\ref{e15}), (\ref{e16}), (\ref{e17}), the relevant terms that violate this symmetry are only the $U$ and $V$ terms and the constraint remains the same. We can then continue and construct all the equations of motion as before.

The easiest way to summarize the result of such analisi is to consider the effective action in the Einstein frame, as we did in the previous case where we did not have gauge fields.
Now, for the case where gauge fields are included in the way described by (\ref{finalaction}), 
all the equations of motion in the Einstein frame will be correctly described by 
\begin{equation}\label{finalbagmodel}
S_{eff}=\int\sqrt{-\bar{g}}d^{4}x\left[-\frac{1}{\kappa}\bar{R}(\bar{g})
+p\left(\phi,R,X,F_{\mu \nu }^{a}, \bar{g}^{\alpha\beta}\right)\right] 
\end{equation}

\begin{equation}\label{newp}
 p = \frac{\chi}{\chi -2 \kappa \epsilon R} \left[X
+\frac{N}{2}\sqrt{-F_{\mu \nu }^{a}F^{a}_{\alpha\beta}\bar{g}^{\mu\alpha}\bar{g}^{\nu\beta}}\right] - \frac{1}{4}F_{\mu \nu }^{a}F^{a}_{\alpha\beta}\bar{g}^{\mu\alpha}\bar{g}^{\nu\beta} 
 - V_{eff}
\end{equation}

\begin{equation}\label{newV}
 V_{eff}  = \frac{\epsilon R^{2} + U}{(\chi -2 \kappa \epsilon R)^{2} }
\end{equation}

where it is understood that,
\begin{equation}\label{chi3}
\chi = \frac{2U(\phi)}{M+V(\phi)}.
\end{equation}

We have again two possible formulations concerning $R$:
Notice first that $\bar{R}$ and $R$ are different objects, the $\bar{R}$ is the Riemannian curvature scalar in the Einstein frame,
while $R$ is a different object. This $R$ will be treated in two different ways:

1. First order formalism for $R$. Here $R$ is a lagrangian variable, determined as follows,  $R$ that appear in the expression above for $p$ can be obtained from the variation of the pressure functional action above with respect to $R$, this gives exactly the expression for $R$ that can be solved for by using the equations of motion in the original frame (and then reexpresing the result in terms of the tilde metric), in terms of $X, \phi$, etc. 

2. Second order formalism for $R$. $R$ that appear in the action above is exactly the expression for $R$ which can be solved from the equations of motion in terms of $X, \phi$, etc. Once again, the second order formalism can be obtained from the first order formalism by solving algebraically R from the eq. obtained by variation of $R$ , and inserting back into the action.
Now $R$ is given by
\begin{equation}\label{newR}
R = \frac{-\kappa (V+M) +\kappa \chi \left( X +\frac{N}{2}\sqrt{-F_{\mu \nu }^{a}F^{a}_{\alpha\beta}\bar{g}^{\mu\alpha}\bar{g}^{\nu\beta}}\right)}
{1 + 2\kappa ^{2} \epsilon \left(X +\frac{N}{2}\sqrt{-F_{\mu \nu }^{a}F^{a}_{\alpha\beta}\bar{g}^{\mu\alpha}\bar{g}^{\nu\beta}}\right) }
\end{equation}

\section{Regular gauge field dynamics inside the bags}
From (\ref{finalbagmodel}), (\ref{newp}) and (\ref{chi3}), we see that the $N$ term, responsible for the confining gauge dynamics, gets dressed in the Einstein frame effective action by the factor $\frac{\chi}{\chi -2 \kappa \epsilon R}$, we will have to check also whether $V_{eff}$  contributes to the gauge field equations of motion.

As we consider regions inside the bags, where $\phi \rightarrow -\infty$, we see  that $\chi$ as given by (\ref{chi3}), approaches zero for 
$M \neq 0$, for the case therefore $\epsilon \neq 0$
the $N$ term inside the bags dissapears. Notice that if we had not introduced the curvature squared term (i.e. if $\epsilon = 0$) this effect would be absent.

In this same limit and with the same conditions, using only that as $\phi \rightarrow -\infty$, $U \rightarrow 0$ and $\chi \rightarrow 0 $, we see that still, in the more complicated theory with gauge fields the same bag constant  $V_{eff} \rightarrow  \frac{1}{4\epsilon \kappa^{2}}$ is obtained, so $V_{eff}$ does not contribute to the gauge field equations of motion, but does provide the Bag constant.

 In the limit $\phi \rightarrow -\infty$, the only term providing gauge field dynamics is the standard term 
$-\frac{1}{4}F_{\mu \nu }^{a}F^{a}_{\alpha\beta}\bar{g}^{\mu\alpha}\bar{g}^{\nu\beta} $.

\section{Confining gauge field effective action outside the bags}
 We are going to assume $M>0$, so to keep $\chi$ positive and finite everywhere and take now the opposite limit, $\phi \rightarrow +\infty$ .
 Furthermore, the choice $M>0$ pushes the scalar field outside the bag to large values of $\phi$, since the absolute minimum of the effective potential is found for such values.
 In this case,

\begin{equation}\label{Vlimit}
 V_{eff}   \rightarrow C+4B \left[\,X
+\frac{N}{2}\sqrt{-F_{\mu \nu }^{a}F^{a}_{\alpha\beta}\bar{g}^{\mu\alpha}\bar{g}^{\nu\beta}}\ \right]^2
\end{equation}

and
 
 $ \frac{\chi}{\chi -2 \kappa \epsilon R}\left[\,X
+\frac{N}{2}\sqrt{-F_{\mu \nu }^{a}F^{a}_{\alpha\beta}\bar{g}^{\mu\alpha}\bar{g}^{\nu\beta}}\ \right]  \rightarrow $ 

\begin{equation}
A\left[\,1+2\kappa^2 \epsilon \left[\, X
+\frac{N}{2}\sqrt{-F_{\mu \nu }^{a}F^{a}_{\alpha\beta}\bar{g}^{\mu\alpha}\bar{g}^{\nu\beta}}\ \right]\ \right] \left[\,X
+\frac{N}{2}\sqrt{-F_{\mu \nu }^{a}F^{a}_{\alpha\beta}\bar{g}^{\mu\alpha}\bar{g}^{\nu\beta}}\ \right]
\end{equation}

where the constants $A$, $B$ and $C$ are given by,
\begin{eqnarray}\label{A}
A &=& \frac{f_2}{f_2 + \kappa^2\epsilon f_1^2}\,,\\
B &=& \frac{\epsilon\kappa^2}{4(1+\kappa^2\epsilon f_1^2/f_2)} = \frac{\epsilon \kappa^2}{4}\,A \,,\label{B} \\
C &=& \frac{f_1^2}{4\,f_2(1+\kappa^2\epsilon f_1^2/f_2)} =
\frac{f_1^2}{4f_2}\,A\,\label{C}.
\end{eqnarray}

Therefore, the resulting dynamics outside the bag, for $\phi \rightarrow +\infty$ will be described by the effective action

\begin{equation}\label{outsidebag}
S_{eff, out}=\int\sqrt{-\bar{g}}d^{4}x\left[-\frac{1}{\kappa}\bar{R}(\bar{g})
+p_{out}\left(\phi,X, F\right)\right] 
\end{equation}
where

\[ \nonumber
p_{out}\left(\phi,X,F\right) = AX+ A \frac{N}{2}\sqrt{-F_{\mu \nu }^{a}F^{a}_{\alpha\beta}\bar{g}^{\mu\alpha}\bar{g}^{\nu\beta}} 
- (1 -4BN^2+ 2N^2\epsilon \kappa^2A)\frac{1}{4}F_{\mu \nu }^{a}F^{a}_{\alpha\beta}\bar{g}^{\mu\alpha}\bar{g}^{\nu\beta}
\]
\begin{equation}\label{poutsidebag1}
+(2AN\epsilon \kappa^2-4BN) X\sqrt{-F_{\mu \nu }^{a}F^{a}_{\alpha\beta}\bar{g}^{\mu\alpha}\bar{g}^{\nu\beta}} +(2A\epsilon \kappa^2-4B)X^{2} -C
\end{equation}

or expressing $B$ in terms of $A$,

\[ \nonumber
p_{out}\left(\phi,X, F\right) = AX+ A \frac{N}{2}\sqrt{-F_{\mu \nu }^{a}F^{a}_{\alpha\beta}\bar{g}^{\mu\alpha}\bar{g}^{\nu\beta}} 
- (1 + N^2\epsilon \kappa^2A)\frac{1}{4}F_{\mu \nu }^{a}F^{a}_{\alpha\beta}\bar{g}^{\mu\alpha}\bar{g}^{\nu\beta}
\]
\begin{equation}\label{poutsidebag2}
+ AN\epsilon \kappa^2 X\sqrt{-F_{\mu \nu }^{a}F^{a}_{\alpha\beta}\bar{g}^{\mu\alpha}\bar{g}^{\nu\beta}} + A\epsilon \kappa^2X^{2} -C
\end{equation}

\section{Decoupling Gravity and the Flat spacetime bag model type action, The bag constant as a coupling constant}
We will discuss the confinement dynamics outside the bags, but to make matters simple and also, most relevant to applications to strong interaction physics, we consider the limit where gravity is decoupled.

As we mentioned before in connection with the bag constant,  we can take the limit $\kappa \rightarrow 0$ and still retail a meaningfull bag constant, if the bag constant is kept  fixed, which implies that 
$\epsilon \kappa ^{2}$ is kept fixed. 

We can see however now that this is true not just for the bag constant, but for the full theory. After we solve for $R$ and then reintroduce this into the action, we can see that in the matter
part of the effective action in the Einstein frame,  $\kappa$ does  not enter by itself, in the matter part $\kappa$ enters only in the combination
$\epsilon \kappa ^{2}$, which is basically the inverse of the bag constant.

The limit $\kappa \rightarrow 0$,  if the bag constant is kept  fixed, which implies that 
$\epsilon \kappa ^{2}$ is kept fixed, decouples the metric of the Einstein frame, which represents "gravity". The combination 
$\epsilon \kappa ^{2}$ remains here as a basic coupling constant in diverse parts on the effective action, that is the Bag constant is a fundamental constant of the effective action.

We see from (\ref{poutsidebag2}) that basically we recover, for the regions of large $\phi$, the square root confinement model, slightly altered by the presence of the scalar field, however in the infrared, when looking at configurations of constant field strength squared, the solutions for the scalar field decay as $1/r$ in flat space, not affecting the infrared behavior of the gauge field dynamics. The square root confinement terms reappear outside the bags, while being absent inside. 

The discussion of the case $N<0$ appears to be the same as that studied before in the "naive" square root gauge theory (that is, not embedded in the TMT model), except for the existence of bag regions where the gauge dynamics recovers its standard dynamics. Outside those bags, we will recover even at the classical level, the Cornell confining potential demonstrating confining properties between charged bags in principle, or there could be just one bag altogether neutral and no gauge fields outside, but the possibility of charged bags exist here.

The case $N>0$ gives also a picture of confinement, but if we work in the naive square root model, as we discussed before, we find in the infrared region a that the field strength must be opposite to itself. In fact assuming that $\phi \rightarrow \infty$, or if $\phi$ is just very big, we obtain now, in the square root theory embedded in TMT, that

\begin{equation}\label{Foutsidebag}
{F^{a\mu \nu } = - \frac{NA}{1+\epsilon \kappa^2 A N^2} \frac{{F^{a\mu \nu } }}{{\sqrt { - F_{\alpha\beta }^b F^{b\alpha \beta } } }}}.
\end{equation}

So again, we find then in the infrared region that the field strength must be opposite to itself. However, for the square root theory embedded in TMT there is a way out of this, since the scalar field $\phi$ does not have to stay in the region of high values, if this means a dynamical inconsistency,
it can return to the region of very negative values and solve this problem. Also, the field strenghts can dissapear before we in fact get to the
 $\phi \rightarrow \infty$ region and this also provides a resolution of the problem .
 
In the context, it means that once we if we go far enough out ot the bag,  the field strength has to either dissapear before (\ref{Foutsidebag})
is applicable avoiding therefore the problem, or alternatively, enter another bag region. This in effect appears to reproduce effectively the MIT bag model boundary conditions, but not as a sharp behavior, but rather, smoothed out over some region.

So we see that the case that most closely resembles the MIT bag model is the case with $N>0$.

\section{Elimination of Massless Scalar Field dynamics other than the dilaton inside the Bags}
If we introduce an additional massless scalar into the theory, then the kinetic term can be incorporated in a conformally invariant way
by coupling it to the measure $\Phi$, as we have done with the square root term of the gauge field squared. Going then to the Einstein frame effective action, we will encounter then exactly the same phenomenon, : as $\phi \rightarrow -\infty$ this term will be supressed.
This is a desirable result, since the Goldstone bosons in some versions of the Bag model, are restricted from entering the bag. Here this result is very natural.

Notice that this does not apply to the dilaton itself. This is because although the dilaton kinetic term is also multiplied by the factor $\chi$
which as $\phi \rightarrow -\infty$ which gives us $\chi \rightarrow 2U/M$, we can define a new scalar field which is for example 
$\Psi =\int \sqrt{2U/M} d\phi $ which will not have a vanishing kinetic term in this limit. Such procedure is not possible for another scalar field different from the dilaton, since in this case there will be no change of variable that will allow to convert the kinetic term into a canonical form.

The question of whether scalar fields, in particular pions, are allowed inside the MIT bag model is a crucial one. It is at the heart of heated discussions on how to formulate the best "chiral bag model". Chiral bag models, see \cite{chiral1},\cite{chiral2},\cite{chiral3}, 
\cite{chiral4},
\cite{chiral5}, \cite{chiral6}, \cite{chiral7},\cite{chiral8},\cite{chiral9}, \cite{chiral10}, \cite{chiral11}, \cite{chiral12}. While in 
refs. \cite{chiral2}, \cite{chiral3}, \cite{chiral4},
\cite{chiral5}, \cite{chiral10}, \cite{chiral11}, the pion is excluded from the bag, but in the "cloudy" version, \cite{chiral1}, \cite{chiral6}, \cite{chiral7},\cite{chiral8}, it is the opposite,  the pion is allowed inside the bag. There is also a middle way, where the pion can enter a part of the bag but not all of it \cite{chiral12}.

This question can be treated in terms of symmetries of the action. As we have seen, the "confinement term", which is the square root of the field strength squared and the kinetic tern of a scalar field must be multiplied both by the same measure of integration $\Phi$, if we insist on these two terms having the local conformal invariance (\ref{e14}), (\ref{e15}), (\ref{e16}), (\ref{e17}). The transition to the unconfined phase inside the bag and the elimination of the scalar field (different from the dilaton)  kinetic term takes place simultaneously due to this basic geometric feature. That is in this picture the version of the chiral bag model where the pion does not enter the confinement region is favored. 
\section{Discussion and Conclusions}
The Two Measures Theory has been studied in the context of cosmolgy because it rearranges the interactions and correlates these to vacuum energies, providing among other things a new approach to the cosmological constant problem.

The strong interactions, in particular viewed from the context of the bag model present us with its own "cosmological ccnstant problem",
which is why in the regions with high vacuum energy (inside the bags)) we obtain free gauge field dynamics, while outside the bags a sort of confinement has to set in.

We saw in this paper that softly broken conformally invariant down to global scale invariance, which is itself spontaneously broken is a very
good tool to approach this problem. The use of an additional measure in particular, allows us to mantain conformal invariance and introduce terms which couple to the new measure and transform different than the terms that couple to $\sqrt{-g}$. Introducing soft breaking of conformal symmetry to global scale invariance and after s.s.b of scale invariance, the confinement dynamics, introduced in a conformally invariant way through a square root of gauge field squared term coupled to the measure $\Phi$ is screened in the regions of high energy density while it becomes "alive" in the regions of low vacuum energy, which repreduces, at a qualitative level the MIT bag model behavior. The square root gauge theory approach to confinement appears then to merge with the bag model approach to confinement, they are not different approaches, but it appears, must be used in conjunction to obtain the most physically appealing picture.

In the limit where gravity is decoupled, we can obtain a finite bag constant and a well defined flat spacetime theory. Different kinetic terms in the action "remember" however the way they coupled to the metric and therefore (after demanding conformal invariance 
(\ref{e14}), (\ref{e15}), (\ref{e16}), (\ref{e17}) with what measure they are coupled. This requirement implies that scalars are also not allowed in the confinement region, favoring a special class of chiral bag models.

Concerning the specifics of the confinement outside the bags, these details depend really on the sign for $N$. If $N>0$, gauge fields can exist outside bags, where there is a linear confinement between two widely separated charged bags, this departs a bit from the standard MIT bag model (because bags can be charged). 

For $N<0$, we find that after we go out from the region where the bag lives, in the limit $\phi \rightarrow \infty$, the field strength is opposite to itself, which means we cannot get to this asymptotic region with a non vanishing field strength. 
The case $N<0$ appears to be more closely related to the MIT bag model, reproducing the MIT bag model boundary condions that implement the vanishing of the field strengths outside the bag. Here it is not a boundary condition but rather a dynamical process that leads effectively to a similar effect.

In a future publication we will address more complete models, including quark fields. 

Also the study of the theory at finite temperature
and the possibility of studying a deconfinement phase transition seems possible here. Indeed, at zero temperature, we have the deconfined false vacuum that exists for $\phi \rightarrow -\infty$ and a true confining vacuum at $\phi \rightarrow \infty$, but at finite temperature, the false vacuum state that exists for $\phi \rightarrow -\infty$ can get stabilized. This can be seen using the canonically normalized field $\Psi =\int \sqrt{2U/M} d\phi $ mentioned before and then appying standard field theory techniques at finite temperature\cite{highT}. 

Finally, it is interesting that the model is defined by introducing gravity, but then turning it down is possible and non trivial results remain, provided the product $\epsilon \kappa^2$ is kept fixed. It looks like gravity is required at least as a regulator to formulate the theory properly. An interesting subject to study could be whether this model can be obtained from regular QCD using standard field theory techniques without reference to gravity. Going on the opposite direction, the model could be also the natural framework for considering gravitational modifications of the confinement dynamics.

\section{Acknowledgements}
I would like to thank Pedro Labrana, Ramon Herrera and Sergio del Campo for many interesting conversations 
relevant to the research carried out here and for collaboration on a related model in the context of cosmology \cite{friends}.
Also I would like to thank Alexander Kaganovich for many discussions and collaboration on TMT and Igor Korover, Patricio Gaete and Euro Spallucci for discussions and collaboration on the square root gauge theory approach to confinement. Finally, I would like to thank Ramy Brustein, Aharon Davidson, Michael Gedalin, Larry Horwitz, David Owen and Moshe Moshe for additional discussions on the square root gauge theory approach of confinement.

\break

\end{document}